\documentclass[english,aps,prl,superscriptaddress,reprint]{revtex4-1}
\usepackage{graphicx}
\usepackage{subfigure}
\usepackage{bm}
\usepackage{amsmath}
\usepackage{tabularx}
\usepackage{array}
\usepackage{multirow}
\usepackage{float}
\usepackage{hyperref}

\begin{document}

\title{Coherence times of precisely depth controlled NV centers in diamond}

\author{Junfeng Wang}
\affiliation{Hefei National Laboratory for Physical Science at Microscale, and
Department of Physics, University of Science and Technology of China,
Hefei, Anhui, 230026, P. R. China}
\author{ Wenlong Zhang}
\affiliation{Hefei National Laboratory for Physical Science at Microscale, and
Department of Physics, University of Science and Technology of China,
Hefei, Anhui, 230026, P. R. China}
\author{Jian Zhang}
\affiliation{Hefei National Laboratory for Physical Science at Microscale, and
Department of Physics, University of Science and Technology of China,
Hefei, Anhui, 230026, P. R. China}
\author{Jie You}
\affiliation{Key lab of Quantum Information, CAS, University of Science and Technology
of China, Hefei, Anhui, 230026, P. R. China}
\author{Yan Li}
\affiliation{Key lab of Quantum Information, CAS, University of Science and Technology
of China, Hefei, Anhui, 230026, P. R. China}
\author{Guoping Guo}
\affiliation{Key lab of Quantum Information, CAS, University of Science and Technology
of China, Hefei, Anhui, 230026, P. R. China}
\author{Fupan Feng}
\affiliation{Hefei National Laboratory for Physical Science at Microscale, and
Department of Physics, University of Science and Technology of China,
Hefei, Anhui, 230026, P. R. China}
\author{Xuerui Song}
\affiliation{Hefei National Laboratory for Physical Science at Microscale, and
Department of Physics, University of Science and Technology of China,
Hefei, Anhui, 230026, P. R. China}
\author{Liren Lou}
\affiliation{Hefei National Laboratory for Physical Science at Microscale, and
Department of Physics, University of Science and Technology of China,
Hefei, Anhui, 230026, P. R. China}
\author{Wei Zhu}
\affiliation{Hefei National Laboratory for Physical Science at Microscale, and
Department of Physics, University of Science and Technology of China,
Hefei, Anhui, 230026, P. R. China}
\author{Guanzhong Wang}\email{gzwang@ustc.edu.cn}
\affiliation{Hefei National Laboratory for Physical Science at Microscale, and
Department of Physics, University of Science and Technology of China,
Hefei, Anhui, 230026, P. R. China}

\begin{abstract}
We investigated the depth dependence of coherence times of nitrogen-vacancy (NV) centers through precisely depth controlling by a moderately oxidative at 580
$^{\circ}$C in air. By successive nanoscale etching, NV centers could be brought close to the diamond surface step by step, which enable us 
to trace the evolution of the number of NV centers remained in the chip and to study the depth dependence of coherence times of NV centers with the diamond 
etching. Our results showed that the coherence times of NV centers declined rapidly with the depth reduction in their last about 22 nm before they finally disappeared, revealing a critical depth for the influence of rapid fluctuating surface spin bath. By monitoring the coherence time variation with depth, 
we could make a shallow NV center with long coherence time for detecting external spins with high sensitivity.
\end{abstract}

\maketitle
In recent years, shallow NV center has attracted increasing attention
owing to its applications in nanoscale spin detection\cite{key-1,key-2,key-3,key-4,key-5,key-6,key-7}
and surface spin noise investigation\cite{key-8,key-9,key-10,key-11}.
Since the minimum detectable magnetic dipole moment scales as $\delta\mu \propto r^{3}/\sqrt{T_{2}}$
\cite{key-9,key-12,key-13}, where r is the NV-target spin distance
and $T_{2}$ is the coherence times of the NV center, the nanoscale
control of the NV center depth and the investigation of depth dependence
of NV center coherence time are important.

Recently, two methods, plasma etching \cite{key-14} and oxidative
etching \cite{key-13,key-15}, have been developed to control the
depth of the NV center. Cui et al. found that the conventional oxygen
plasma etching had low damage to diamond \cite{key-14}. For oxidative
etching, Kim et al. demonstrated that the diamond could be etched
at the temperature, ranging from 550 $^{\circ}$C to 620 $^{\circ}$C, in pure oxygen gas
\cite{key-13}. Loretz et al. reported that the oxidative etching
rate for diamond was about 10 $nm/h$ at 650 $^{\circ}$C in air and by using
this method they realized a 1.9-nm-deep NV center \cite{key-15}.
However, the etching rate was not short enough to precisely control the depth of NV center .

In this work, we performed oxidative etching in air on diamond at a reduced oxidative temperature
580 $^{\circ}$C, obtaining about 1.1 nm/h etching rate. By successive nanoscale etching, NV centers
 could be brought close to the diamond surface step by step, which enable us to investigate the depth
 dependence of coherence times of NV centers. The coherence time of NV center declined rapidly with the
 depth reduction in their last about 22 nm before they finally disappeared, which was attributed to the influence 
of the rapid fluctuating surface spin bath. By monitoring the coherence time variation with depth, 
we could make a shallow NV center with long coherence time for detecting external spins with high sensitivity\cite{key-5,key-7,key-15,key-16}. 

\begin{figure}
\includegraphics{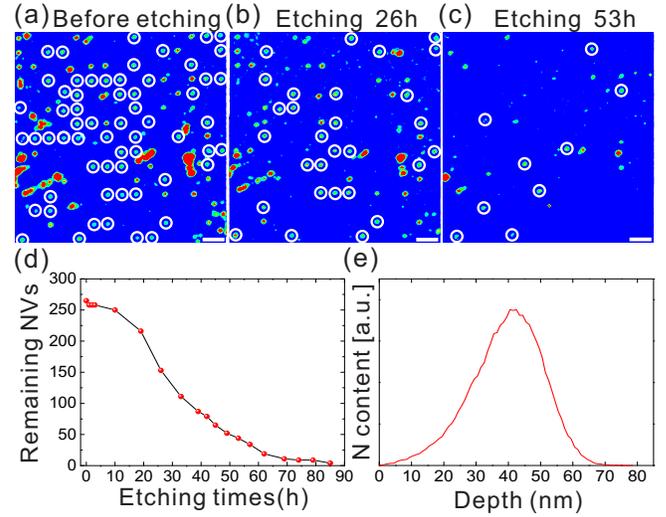}

\protect\caption{Fluorescence images of the remaining NV centers in the diamond chip
after different oxidative etching times. (a) Image of a representative
area of the chip before oxidative etching. (b) and (c) The same area
after oxidative etching 26 h and 53 h, respectively. The circled single
NV centers in the (a) ((b) and (c)) were used for tracing. All the
scale bars in (a) ((b), (c)) are 5 $\mu m$. (d) The total number
of the remaining NV centers in the selected area on the diamond chip
after different oxidative etching times. (e) The SRIM simulations
of the depth profile of the implanted nitrogen atoms at an energy
of 30 keV. }
\end{figure}

\begin{figure*}
\includegraphics{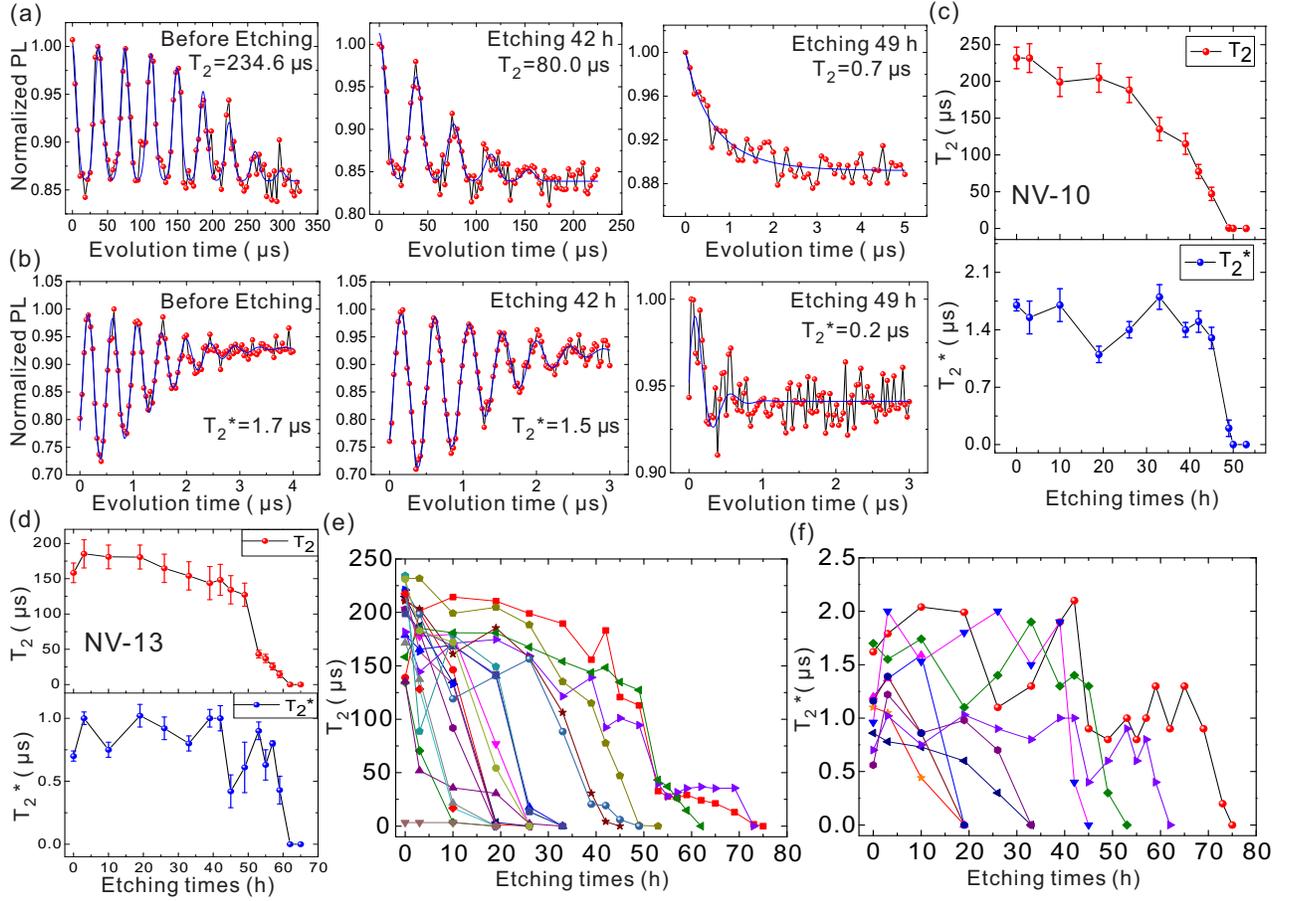}

\caption{Coherence times measurements of the NV centers after different oxidative
etching times. (a) and (b) Spin echo and Ramsey measurements, respectively,
of NV-10 after three different oxidative etching times. The blue lines
were the fits of the $I_{PL}$. (c) and (d) The spin echo and Ramsey
measurements of NV-10 and NV-13 after different oxidative etching
times, respectively. (e) Coherence times vs. etching time  derived
from spin echo measurements for all the 20 single NV centers. (f)
Coherence times vs. etching time derived from Ramsey measurements
 for 9 representative single NV centers.}
\end{figure*}

\begin{figure*}
\includegraphics{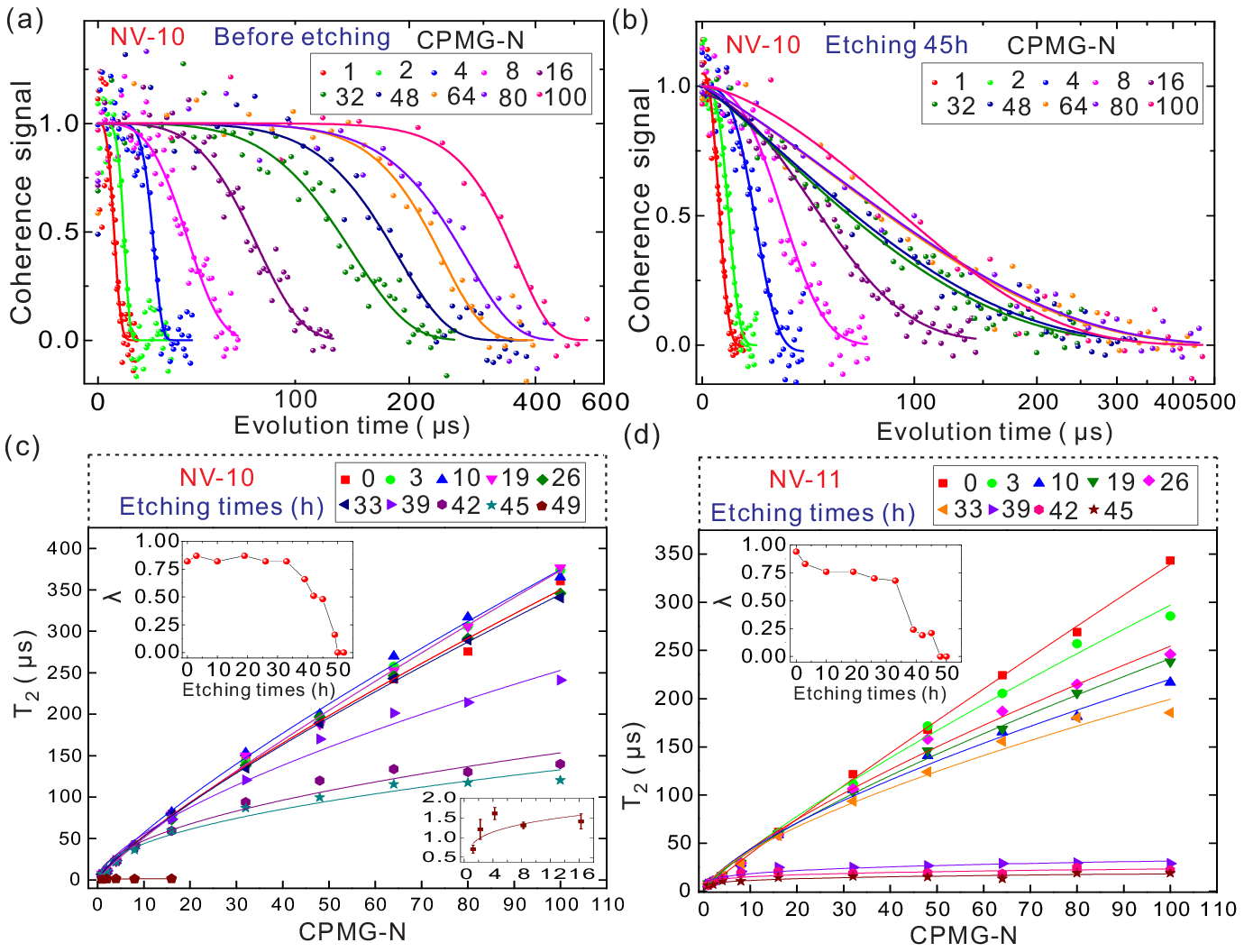}

\caption{Dynamical decoupling 
measurements after different oxidative
 etching times. (a) and (b) The coherence
 decay curves of NV-10 in CPMG-N with the
 $ \pi$ pulse number N from 1 to 100 for 
the sample before etching and with etching
 45 h, respectively. The solid lines were 
the fits to the data using the exponentials 
$e^{-(t/{T_{2}})^{n}}$,where $T_{2}$ is
 the coherence times. (c) The coherence times $T_{2}$
of the CPMG of NV-10 as functions 
of N for the chip etched for different
times. The inset (bottom right) was the
 zoom of the CPMG-N (N from
1 to 16) results for the chip after 
etching 49h. (d) The coherence
times $T_{2}$ of the CPMG of NV-13 
as functions of N after different
oxidative etching times. The solid 
lines in the (c) and (d) were the
fits to the data using the power-law
 $T_{2,CPMG-N}/T_{2,echo}=N^{\lambda}$
and the decoupling efficiency $\lambda$ as
 a function of the etching
time for both cases were shown
 in the inset (top left) of
 the corresponding figures, respectively.}
\end{figure*}

A $2\times 2 \times 0.5 $ mm$^{3}$ (100) electronic
grade diamond chip from Element Six ([$^{13}C$]=1.1\%, [N]\textless 5 ppb) was used for the experiments. 
At first, a 300-nm-thick polymethyl
methacrylate (PMMA) layer was deposited on the diamond chip surface
\cite{key-17,key-18}. Then a series of 60 nm diameter apertures,
as well as some 10-$\mu m$ -wide strips used as position marks \cite{key-18},
were patterned on the PMMA layer using electron beam lithography.
The NV centers in the chip were generated by implanting 60 keV $^{14}N_{2}^{+}$
molecules with a fluence $ 0.55 \times 10^{11} $
$^{14}N_{2}^{+}/cm^{2}$ through the apertures and strips on the
PMMA layer.\cite{key-17,key-18}. After implantation, the sample was
annealed at 1050 $^{\circ}$C in a vacuum at $2 \times 10^{-5}$
Pa for 2 h to form long spin coherence time NV centers \cite{key-19}.
After oxidation at 430 $^{\circ}$C in air for 2.5 h, the sample was cleaned
in a 1:1:1 boiling mixture of sulfuric, nitric, and perchloric acid
at 200 $^{\circ}$C for 2 hours.

The oxidative etching was performed in air on the diamond chip situated
in a box furnace with its implanted side up. In order to have a slow
etching rate, we used a lower oxidative temperature of 580 $^{\circ}$C. The
diamond chip was put in the furnace when the furnace temperature had
already stabilized at 580 $^{\circ}$C. The process for the furnace temperature
being stabilized at 580 $^{\circ}$C only took about 1 minute. Similarly, when
the etching process was finished, we directly took the diamond chip
out to the air for cool down. We traced the evolution of the number
of single NV centers remaining in the chip with the etching time and
investigated the relationship between the coherence times of the NV
centers and the etching times. Figure 1(a) showed the fluorescence
image of a representative area of the diamond chip before oxidative
etching, with the traced single NV centers marked with circles in
the figure. The larger bright specks in the image were NV center clusters
produced in the implantation due to imperfection of the PMMA templet
layer deposited on the diamond chip \cite{key-18}. After oxidative
etching for 26 h, the image of the same area showed less single NV
centers and the bright specks of the NV center clusters also became
less bright and smaller [Fig. 1(b)]. As shown in Fig. 1(c), after
oxidative etching for 53 h, there were only few single NV centers
remained on the chip and the bright specks became even less bright and
smaller. These results clearly demonstrated that the NV centers originally
situated at different depths of the diamond chip had been removed
successively by oxidative etching \cite{key-13,key-15}.

We traced 265 single NV centers that were initially in the diamond
chip. As shown in Fig. 1(d), the number of the remaining single NV
centers in the chip reduced with the oxidative etching times. We found that when the etching time was less than 10 h, the number of the
remaining NV centers reduced very little. Then, when the etching times
increased from 10 h to about 60 h, the NV centers number reduced rapidly.
After that, the number reduced slowly and finally almost all the NV
centers finally disappeared after 85 h etching. The results showed that the number
of the remaining single NV centers reduced about 70\% after etching
42 h. By comparing the above results with the SRIM simulations for
an implantation nitrogen atom energy of 30 keV [Fig. 1(e)], we
found that the etching depth corresponding to 42 h etching was 46
nm. Therefore, we estimated that the etching rate was about 1.1 nm/h at 580
C in air. 

The slow etching rate made it possible to control the depth of the
NV center at nanoscale. Thus, we could precisely trace the evolution
of the coherence time of a single NV center with its depth, which
reflected the variation in its local environment, from the bulk $^{13}C$
spin bath to the surface spin bath. A set of randomly selected 20
single NV centers [named NV-1 through 20] with their axes parallel
to the external magnetic field (47 G) was used for the tracing. We
performed spin echo measurements for all the 20 single NV centers
before the chip etching, and found that the $T_{2}$ of 19 NV centers
among them were between 134 and 234 $\mu s$, and the other one was
3.3 $\mu s$. The mean value of $T_{2}$ of the 19 single NV centers
was about 194 $\mu s$, which was comparable to that of the native
NV center inside electronic grade diamond \cite{key-20}, implying
that the NVs were initially deep inside the chip and the initial major
spin impurities were $^{13}C$. 

Then, we measured the coherence times of the 20 NV centers after successively etching
for various times step by step. Fig. 2(a) showed three representative results of
the spin echo measurements for NV-10 (which just disappeared after
etching 50 h), after different oxidative etching times. In the first
two figures, we found that the photoluminescence (PL) profile
collapsed and revived at $^{13}C$ Larmor frequency (about 50.3 kHz),
which was due to the incoherent precession of the \textsuperscript{13}C
nuclei spins in our natural $^{13}C$ abundance samples\cite{key-21}.
It was found that for the sample before etching, the $T_{2}$ of NV-10 was 234.6 $\mu s$. And, after an etching of 42 h, the $T_{2}$ declined
to 80 $\mu s$. However, when the etching time reached 49 h, its PL intensity was not stable and was reduced to about half
of the initial value, indicating that the center had been very shallow
and its charge state had become unstable \cite{key-22,key-23}. (The
similar phenomenon was also observed for other NV centers, when they
were shallow enough by long time etching.) Moreover, its $T_{2}$
was reduced to only about 0.7 $\mu s$. The above results revealed
that the NV center became very shallow and suffered a strong influence
of the rapid fluctuating surface spin bath \cite{key-8,key-9,key-10,key-24}.
The $T_{2}$ of NV-10 in the sample etched for various times were
presented in Fig. 2(c) top. This etching time dependence can be converted
to the depth dependence as the depth varied with the etching time.
As shown from the figure, for short etching time (less than about 30
h), the $T_{2}$ decreased slowly with the etching time, then declined
rapidly in its last about 20 h etching before it finally disappeared (i.e.
when its depth was less than about 22 nm). Finally, when the etching
time was longer than 50 h, NV-10 was finally disappeared.

Three representative Ramsey measurements for NV-10 were also shown
in Fig. 2(b), and its $T_{2}^{*}$ for the sample after various etching
times were summarized in Fig. 2(c) bottom. We could see that the $T_{2}^{*}$
had a similar evolution to the $T_{2}$. As another example, Figure
2(d) gave the results of the coherence time measurements for NV-13
(just disappeared after etching 62 h). It is clear that the coherence
time decreased very slowly with the etching time in the initial $\sim$
40 h, while declined rapidly in the last $\sim$ 20 h etching before
it finally disappeared (i.e. when its depth was less than about 22 nm). The
evolution were similar to that of NV-10. Fig. 2(e) summarized the
dependences of $T_{2}$ with the etching time for all the 20 NV centers.
We found that there was a similar dependence for all these
NV centers: the $T_{2}$ of each NV centers decreased rapidly when
the etching time was about 20 h before the centers finally disappeared (i.e.
when their depth became less than about 22 nm). The results of Ramsey
measurements of the 9 representative single NV centers after different
oxidative etching times were shown in Fig. 2(f), which exhibited a
similar variation as that of their $T_{2}$ in Fig. 2(e). The results suggested that,
when the depth decreased to about 22 nm, rapidly fluctuating surface
spins, compared with the slow bulk $^{13}C$ spin bath, began to play
an important role in NV decoherence, leading to a rapid decreasing
in coherence time with a further decrease in centers depth, which
was consistent with the results of the delta doping NV centers \cite{key-9}. 

Dynamical decoupling was important in NV center based high sensitivity
magnetic field \cite{key-25} and temperature \cite{key-18,key-26} sensing.
In particular, the shallow NV centers, combined with dynamical decoupling,
could also be used to detect external spins \cite{key-3,key-5,key-6,key-7}
and investigate the surface spin noise \cite{key-9,key-10,key-11}.
In view of this, we performed CPMG measurements on three arbitrarily
selected NV centers (NV-1, 10, 11) on the chip after different oxidative
etching times. The results were shown in Fig. 3. As shown in Fig.
3(a), before etching, the $T_{2}$ of the CPMG-N of NV-10 were always
increasing with the $\pi$ pulse number N, in particular, the
$T_{2}$ of the CPMG-100 was about 360.4 $\mu s$, about 52 times
longer than the value 6.9 $\mu s$ for the spin echo. However, in
Fig. 3(b), after etching for 45 h, almost all the $T_{2}$ of the
CPMG-N were declined, in particular, the $T_{2}$ of the CPMG-100
was about 120.4 $\mu s$, which was only about one third of the value
before etching. The $T_{2}$ of the CPMG-N as a function of N for
NV-10 with the chip etched for various times were summarized in Fig.
3(c). The decoupling efficiencies $ \lambda $ \cite{key-9} varied
with the etching time were shown in the inset (top left) of Fig. 3(c).
When the etching time was less than about 30 h, the $T_{2}$ of the
CPMG were decreased very slowly, and the decoupling efficiencies $ \lambda $
was nearly a constant of about 0.82. This was consistent to the fact
that the center, for sample etched for 30 h, was still inside the
chip and the main spin bath was a bulk $^{13}C$ bath\cite{key-20}.
Then, they both declined rapidly when its depth decreasing in its
last about 20 h etching before it finally disappeared (i.e. when its depth
was less than about 22 nm), which was similar to the results in Fig.
2(c). The similar phenomena were observed for NV-11 (just disappeared
after etching 48 h) [Fig. 3(d)] and NV-1 (just disappeared after
etching 75 h, and its data was not shown in this paper). The results were consistent
with the fact that when the depth of the NV centers decreased, they
would have a higher coupling strength with surface spin bath that
had a fast fluctuation rate \cite{key-9,key-10}.

In summary, we investigated the depth dependence of coherence times
of NV centers in diamond with the depth controlled using oxidative
etching. We used a lower oxidative temperature of 580 $^{\circ}$C in air to
reduce the etching rate to about 1.1 nm/h. Thus, we can more precisely
control the depth of the centers. On the basis of this, we performed
the spin echo and Ramsey and CPMG measurements on the NV centers,
of which the depth was decreased by successive oxidative etching,
and found that the coherence times of NV centers declined rapidly with the depth reduction in their last about 22 nm etching before they finally disappeared, revealing a critical depth for the influence of rapid fluctuating surface spin bath, which was consistent with the results of the delta doping NV centers. \cite{key-9}.
The obtained results showed how the surface spin influenced the coherence
behavior of NV centers of different depths, paving the way for investigating
surface spins. Particularly, the slow etching method make it possible
to control the depth of the NV center at nanoscale, which can be used
to study the stability of the very shallow NV centers \cite{key-24,key-25,key-27},
to investigate the surface spin noise \cite{key-8,key-9,key-10,key-11},
and for high sensitivity (even for single electron and nuclear) nanoscale
spin detection \cite{key-5,key-7,key-15,key-16}. Moreover, it can
also be used to control the depth of the NV center in the diamond
tip equipped to the atomic force microscope, which can be used in
high sensitivity nanoscale magnetic imaging of the spins \cite{key-28}.

This work was supported by the National Basic Research Program of
China (2013CB921800, 2011CB921400) and the Natural Science Foundation
of China (Grants No. 11374280 and No. 50772110).

\end{document}